\documentclass[9pt,letterpaper,technote]{IEEEtran}
\usepackage[cmex10]{amsmath}
\usepackage{amsfonts,amsthm,bm}
\usepackage{cite,array}
\usepackage{graphicx}% Include figure files
\usepackage{dcolumn}% Align table columns on decimal point
\usepackage{subfig}
\def\ket#1{| #1 \rangle}
\def\bra#1{\langle #1 |}
\def\ip#1#2{\langle #1 | #2 \rangle}

\def\diag{\operatorname{diag}}
\def\dim{\operatorname{dim}}

\def\Tr{\operatorname{Tr}}

\def\C{\mathcal{C}}
\def\H{\mathcal{H}}
\def\M{\mathcal{M}}
\def\N{\mathcal{N}}

\def\T{\mathcal{T}}

\def\su{\mathfrak{su}}

\def\SU{\mathbb{SU}}
\def\UU{\mathbb{U}}

\def\RR{\mathbb{R}}

\def\ONE{\mathbb{I}}
\def\xs{\vec{x}\cdot\vec{\sigma}}
\def\sx{\sigma_x}

\def\sz{\sigma_z}

\newtheorem{example}{Example}[section]
\newtheorem{remark}{Remark}[section]

\newtheorem{theorem}{Theorem}[section]

\newtheorem{lemma}{Lemma}[section]

\newtheorem{definition}{Definition}[section]

\begin{document}
\title{Analysis of Effectiveness of Lyapunov Control for
Non-generic Quantum States}
\author{\IEEEauthorblockN{Xiaoting Wang\IEEEauthorrefmark{1} and
        S. G. Schirmer\IEEEauthorrefmark{1}\IEEEauthorrefmark{2}}\\
\IEEEauthorblockA{\IEEEauthorrefmark{1}
                  Dept of Applied Maths \& Theoretical Physics \\
                  Univ. of Cambridge, Wilberforce Rd, Cambridge, CB3 0WA, UK\\
                  Email: x.wang@damtp.cam.ac.uk, sgs29@cam.ac.uk}}
%\date{\today}
\maketitle

\begin{abstract}
A Lyapunov-based control design for natural trajectory-tracking problems
is analyzed for quantum states where the analysis in the generic case
is not applicable.  Using dynamical systems tools we show almost global
asymptotic stability for stationary target states subject to certain
conditions on the Hamiltonians, and discuss effectiveness of the design
when these conditions are not satisfied.  For pseudo-pure target states
the effectiveness of the design is studied further for both stationary
and non-stationary states using alternative tools.
\end{abstract}

\section{Introduction}
\label{sec:intro}

Recent technological advances have prompted significant interest in
developing the foundations of quantum control theory.  One major concern
is how to design the dynamics to steer the system state to a stationary
target state, or track the natural trajectory of non-stationary states.
Numerous results have been obtained on the theory of Lyapunov-based
control~\cite{Vettori,Ferrante,Grivopoulos,Mirrahimi2004a,Mirrahimi2004b,
Mirrahimi2005,Mirrahimi-implicit,altafini1,altafini2,Mirrahimi-stabilize,
D'Alessandro} and the references in \cite{D'Alessandro}.  Most papers
have used the Hilbert-Schmidt (HS) distance as the most natural Lyapunov
function and in this setting \cite{Mirrahimi2004a, Mirrahimi2004b,
Mirrahimi2005} showed that the target state is asymptotically stable
under a sufficient condition equivalent to controllability of the
linearized system for pure states represented by wavefunctions, although
an additional control had to be added to fix the relative phase of the
state.  An alternative design based on an implicit Lyapunov function was
proposed in \cite{Mirrahimi-implicit} to render (pure) target states
asymptotically stable when the linearized system is not controllable.
The more general case of systems whose states must be represented by
density operators was recently considered in~\cite{altafini1,altafini2},
but our analysis for generic quantum states~\cite{WS-LyaGeneric}, for
instance, showed the dynamical landscape and convergence behavior to be
more complicated than described in~\cite{altafini2}.  The aim of this
TechNote is to extend this analysis to non-generic states, which,
although a set of measure zero, deserve special consideration as they
include important cases such as pseudo-pure states.

Specifically we show that, although the LaSalle invariant set for
non-generic states is much larger than in the generic case, for ideal
systems the target state not only remains isolated and thus locally
asymptotically stable, but a dynamical systems analysis for stationary
target states shows that all other critical manifolds are unstable and
attract at most a measure-zero subset of the state space.  At least for
the stationary target states, these arguments rigorously show that we
indeed have almost global convergence as claimed in \cite{altafini2},
although the set of states \emph{not} attracted to the target state is
significantly larger than claimed in~\cite{altafini2}.  The conditions
for effectiveness of the method for stationary target states are similar
to those for generic states, and we again find that center manifolds
emerge around the target state when these conditions are not satisfied,
but in contrast to the generic case, the emergence of a center manifold
now is conditional on the target state, i.e., the method may still be
effective for some target states even if the ideal Hamiltonian criteria
are not satisfied.  For the special class of pseudo-pure states, we can
further show explicitly that for systems with ideal Hamiltonians, the
method is effective not only for stationary target states but also for
almost all non-stationary ones.  This is similar to the situation for
the generic case but in this special case we can characterize the
non-regular target states explicitly, which shows that they include some
states of special interest such as so-called CAT states and maximally
entangled states.

\section{Lyapunov control and LaSalle invariant set}
\label{sec:basics}

We study the bilinear Hamiltonian control problem for a quantum system
on an $n$-dimensional Hilbert space $\H$.  The state of such a system is
generally represented by a trace-one positive operator $\rho$ on $\H$,
referred to as \emph{density operator}, whose evolution is governed by
the quantum Liouville equation
\begin{equation}
  \dot \rho(t) = -i [ H_0 + f(t)H_1, \rho(t) ],
\end{equation}
where $f(t)$ is an admissible real-valued control, and $H_0$ and $H_1$
are time-independent Hamiltonians corresponding to free evolution and
control interaction terms, respectively.

\begin{definition}
A density operator $\rho$ represents a \emph{pure} state if it is a
rank-one projector, and a \emph{mixed} state otherwise;  $\rho$ is
generic if it has $n$ distinct eigenvalues, and \emph{pseudo-pure}
if its spectrum has only two distinct eigenvalues occurring with
multiplicities $1$ and $n-1$.
\end{definition}

The control problem is to design a control $f(t)$ such that the system
state $\rho(t)$ converges to the target state $\rho_d$ as $t\to\infty$.
For Hamiltonian evolution a necessary condition for $\rho_d$ to be
reachable from an initial state $\rho(0)$ is that both $\rho(0)$ and
$\rho_d$ have the same spectrum, and we shall assume this condition to
be satisfied here.  The set of density operators isospectral with
$\rho_d$ forms a compact manifold $\M$, whose dimension depends on the
spectrum of $\rho_d$.  $\rho_d$ is a stationary state of the system if
and only if $[H_0,\rho_d]=0$.  If $\rho_d$ is not a stationary state
then the control problem becomes a trajectory-tracking problem, and to
be able to apply LaSalle's invariance principle to this case, we
formulate the control dynamics on an extended state space $\M\times\M$:
\begin{subequations}
\begin{align}
 \dot{\rho}(t)   &= -i [H_0+f(\rho,\rho_d)H_1,\rho(t)],\\
 \dot{\rho}_d(t) &= -i [H_0,\rho_d(t)],\\
 f(\rho,\rho_d)  &= \Tr([-iH_1,\rho(t)]\rho_d(t)),
\end{align}
\label{eqn:auto}
\end{subequations}
where the control function $f$ is chosen such that
\begin{equation}
 \textstyle
 V(\rho,\rho_d) = \frac{1}{2}\Tr[(\rho-\rho_d)^2]
                = \Tr(\rho_d^2)-\Tr(\rho\rho_d)
\end{equation}
is non-increasing along any flow, i.e.,
\begin{equation*}
  \textstyle
  \frac{d}{dt}V(\rho,\rho_d)
  = -f(t)\Tr(\rho_d[-iH_1,\rho])= -f^2(t)\le 0.
\end{equation*}
Thus, the dynamical system $(\rho(t),\rho_d(t))\in\M\times\M$ is
autonomous, all solutions are bounded, and $V(\rho_1,\rho_2)$ is a
Lyapunov function on $\M\times\M$. From LaSalle's invariance
principle~\cite{lasalle} we have:

\begin{theorem}
\label{thm:lasalle:2} Any evolution $(\rho(t),\rho_d(t))$ for the
dynamical system (\ref{eqn:auto}) will converge to the LaSalle invariant
set, $E=\{(\rho_1,\rho_2)\in \M \times\M | \dot{V}(\rho(t),\rho_d(t))=0,
(\rho(0),\rho_d(0))=(\rho_1,\rho_2)\}$.
\end{theorem}

The effectiveness of the control in steering the system towards the
target state depends on the asymptotic stability of $\rho_d$.  Local
asymptotic stability is a prerequisite for almost global convergence,
and a necessary and sufficient condition for the former is that the
target state be isolated in $E$.  Hence, we first characterize the
invariant set $E$, which depends on both the Hamiltonians $H_0$, $H_1$
and $\rho_d$.

\begin{definition}
Choose a basis such that $H_0=\diag(a_1,\ldots,a_n)$, which is always
possible as $H_0$ is Hermitian.  Let $H_1=(b_{k\ell})$ and
$\omega_{k\ell}=a_k-a_{\ell}$ be the transition frequencies.  
System~(\ref{eqn:auto}) is \emph{ideal} if
\begin{itemize}
\item[(i)] $H_0$ is strongly regular, i.e.,
           $\omega_{k\ell}\neq \omega_{pq}$ unless $(k,\ell)=(p,q)$.
\item[(ii)] $H_1$ is fully connected, i.e., $b_{k\ell}\neq 0$ except
            (possibly) for $k=\ell$.
\end{itemize}
\end{definition}

As shown in~\cite{WS-LyaGeneric}, we have
\begin{theorem}
\label{thm:lasalle:4} 
The invariant set of (\ref{eqn:auto}) for an ideal system satisfies
$E=\{(\rho_1,\rho_2)\in \M\times\M:[\rho_1,\rho_2]=\diag(c_1,\ldots,c_n)\}$.
\end{theorem}

\section{Stationary target state}
\label{sec:non-generic:general}

For a generic stationary state $\rho_d$, it has been shown that the
LaSalle invariant set $E$ is comprised of $n!$ distinct stationary
states, which coincide with the $n!$ critical points of $V(\rho)=
V(\rho,\rho_d)$ as a Morse function~\cite{WS-LyaGeneric}.  If $\rho_d$
is not generic then $E$ is much larger and the topology becomes quite
complicated.  If $\rho_d$ is stationary then (\ref{eqn:auto}) can be
reduced to an autonomous system on $\M$, and the LaSalle invariant set
to $E=\{\rho :[\rho,\rho_d]=0\}$.  We can also choose a basis such that
$H_0$ and $\rho_d$ are simultaneously diagonal,
$H_0=\diag(a_1,\ldots,a_n)$,
\begin{equation}
  \label{eqn:degenerate}
  \rho_d=\diag(w_1,\ldots,w_1,\ldots,w_k,\ldots,w_k),
\end{equation}
where $\{w_\ell\}_{\ell=1}^k$ are the distinct eigenvalues of $\rho_d$
with multiplicities $n_\ell$ and $\sum_{\ell=1}^k n_\ell=n$.  The state
space in this case is the flag manifold
\begin{equation}
 \M \simeq \UU(n)/ \left[\UU(n_1)\times \ldots \times \UU(n_k)\right]
\end{equation}
of dimension $n_\M=n^2-\sum_{\ell=1}^k n_\ell^2$.

\begin{definition}
$x_0$ is a critical point of $f(x)$ if $\nabla f(x)|_{x=x_0}=0$; it
is hyperbolic, if the Hessian at the critical point is nonsingular.
$f$ is a Morse function if all of the its critical points are
hyperbolic.
\end{definition}

For a given $\rho_d$, we can investigate the critical points of the
Lyapunov function $V(\rho)=V(\rho,\rho_d)$ on $\M$.

\begin{lemma}
[proof in~\cite{WS-LyaGeneric}] \label{lemma:crit:1} For a given
stationary target state $\rho_d$ the critical points of
$V(\rho)=V(\rho,\rho_d)$ on $\M$ are such that $[\rho,\rho_d]=0$.
\end{lemma}

If $\rho_d$ is generic, there are $n!$ critical points of $V$ in
total; if $\rho_d$ is non-generic, there are more critical points,
forming different critical manifolds (see also~\cite{Wu}).

\begin{example}
\label{eg:1} For a three-level system with
$\rho_d=\diag(\frac{1}{4},\frac{1}{4},\frac{1}{2})$ the set of
$\rho\in\M-\{\rho_d\}$ that commute with $\rho_d$ has the form
\begin{align*}
 \rho_0=
  \begin{pmatrix}
  w_{11}   & w_{12} & 0 \\
  w_{12}^* & w_{22} & 0 \\
  0        & 0      & \frac{1}{4}
  \end{pmatrix},
\end{align*}
which forms a manifold $\M_0$ isomorphic to the Bloch sphere, on
which $V$ assumes its global maximum, while $\rho=\rho_d$
corresponds to minimum of $V$.
\end{example}

\begin{theorem}
\label{thm:degenerate:1} Let $\rho_d$ be a given stationary target
state.  Then $\rho=\rho_d$ is an isolated hyperbolic critical point of
$V(\rho)$ corresponding to the global minimum and there are no other
local or global minima.
\end{theorem}

\begin{IEEEproof}
$V(\rho)$ clearly assumes its global minimum for $\rho=\rho_d$.  To show
$\rho$ is a hyperbolic minimum of $V$ it suffices to show that it is a
hyperbolic maximum of $J(\rho)=\Tr(\rho\rho_d)$.  Choose a basis so that
$\rho_d$ is diagonal and let $\rho=\rho_0$ be a critical point.  Any
point in a neighborhood of $\rho_0$ can be written as
$\rho=e^{\xs}\rho_0e^{-\xs}$, where $\vec{\sigma}=\{\lambda_{k\ell},
\bar{\lambda}_{k\ell},\lambda_k\}$ is the basis of the Lie algebra
$\su(n)$ defined in the Appendix.  Substituting this into $J$ gives
$J=\Tr(e^{\xs}\rho_0e^{-\xs}\rho_d)$.  To show that the critical point
$\rho_0=\rho_d$ is a maximum of $J$, we need to find $n_\M=\dim\M$
independent directions in which $J$ is a local maximum.  If we choose
curves through $\rho_0$ with $\xs=\lambda_{k\ell}t$ then
\begin{align*}
J = & \Tr (\rho_0\rho_d) + t^2
      \{-\Tr(\rho_0\lambda_{k\ell}^\dagger\rho_d\lambda_{k\ell})\\
    & +\mbox{$\frac{1}{2}$} \Tr(\rho_0\rho_d\lambda_{k\ell}^2)
      +\mbox{$\frac{1}{2}$} \Tr(\rho_0 \lambda_{k\ell}^2\rho_d)\}
      + \Theta(|t|^3).
\end{align*}
The conjugate action of $\lambda_{k\ell}$ on $\rho=\rho_d$ on the
$(k,\ell)$ subspace swaps the $k$-th and $\ell$-th diagonal
elements.  If we choose the curve with $\xs=\bar\lambda_{k\ell}$, we
get a similar result.  Hence the number of swaps that decrease the
value of $J$ is
\begin{align*}
 2 \left(n_1 \sum_{\ell=2}^k n_\ell
    +\cdots+n_{k-1} n_k \right)= n^2-\sum_{\ell=1}^k n_\ell^2=\dim\M.
\end{align*}
Thus $\rho=\rho_d$ is a hyperbolic point of $J$, which is
necessarily isolated. At other critical points, it is easy to see
that there always exists some swap $\lambda_{k\ell}$ that increases
the value of $J$, and hence all the other critical points cannot be
local maxima of $J$, or local minima of $V$.
\end{IEEEproof}

\begin{remark}
For a given $\rho_d$, the critical points of
$V(\rho)=V(\rho,\rho_d)$ form a finite number of isolated critical
manifolds with $\rho_d$ as an isolated hyperbolic minimum.  When
$\rho_d$ becomes generic the critical manifolds are reduced to $n!$
isolated hyperbolic critical points.
\end{remark}

From Lemma~\ref{lemma:crit:1}, we can see that for ideal Hamiltonian and
stationary $\rho_d$, the LaSalle invariant set $E$ coincides with the
critical points of $V(\rho)=V(\rho,\rho_d)$, and there are
$p=\frac{n!}{n_1!\cdots n_k!}$ stationary solutions including $\rho_d$,
either isolated or located on an isolated manifold in $E$.  In
particular, since $\rho_d$ is isolated in $E$, the LaSalle invariance
principle guarantees its local asymptotic stability, but if we want to
investigate the asymptotic stability and convergence properties on a
large scale, we must analyze the eigenvalues of the linearized dynamics
at the other stationary states.

\begin{theorem}
\label{thm:dengerate:2} For ideal systems the stationary state $\rho_d$
is a hyperbolic sink of the dynamical system~(\ref{eqn:auto}) and thus
locally asymptotically stable, while all other stationary solutions have
unstable manifolds.  Hence, $\rho_d$ is almost globally asymptotically
stable.
\end{theorem}

\begin{IEEEproof}
In order to analyze the stability at the stationary state we reformulate
the dynamical system~(\ref{eqn:auto}) in the Bloch representation
\begin{subequations}
\begin{align}
\dot {\vec{s}}(t)   &= (A_0+f(\vec{s},\vec{s}_d)A_1)\vec{s}(t)\\
\dot {\vec{s}}_d(t) &= A_0\vec{s}_d(t)\\
f(\vec{s},\vec{s}_d)&= \vec{s_d}^TA_1\vec{s},
\end{align}
\end{subequations}
where $\vec{s}=(s_k)_{k=1}^{n^2-1}$ with $s_k=\Tr(\sigma_k \rho)$, and
$A_0$ and $A_1$ are the anti-symmetric matrices
\begin{subequations}
\begin{align}
A_0 (k,k')  &= \Tr(iH_0[\sigma_k,\sigma_{k'}]),\\
A_1 (k,k')  &= \Tr(iH_1[\sigma_k,\sigma_{k'}]),
\end{align}
\end{subequations}
$\sigma_k$ being the elements of the basis of $\su(n)$ defined above.
For stationary $\rho_d$, the dynamics is reduced to 
\begin{subequations}
\label{eq:sys_real}
\begin{align}
\dot {\vec{s}}(t) &= (A_0+f(\vec{s})A_1)\vec{s}(t)\\
        f(\vec{s})&= \vec{s_d}^TA_1\vec{s}.
\end{align}
\end{subequations}

The linearized system near the critical point $\vec{s}_0$ is
\begin{equation}
\label{eqn:linear}
  \dot {\vec{s}}= D_f(\vec{s}_0)\cdot (\vec{s}-\vec{s}_0),
\end{equation}
where $D_f(\vec{s}_0)=A_0+A_1 \vec{s}_0\cdot \vec{s_d}^T A_1$ is a
linear map defined on $\RR^{n^2-1}$.  (The map is defined on
$\RR^{n^2-1}$ as we neglect the $s_0$ coordinate of $\rho$, which is
constant due to trace preservation.) In order to show $\vec{s}_d$ is
hyperbolic, it suffices to show that there are $n_{\M}$ eigenvalues
with nonzero real parts, corresponding to $n_{\M}$ eigenvectors in the
tangent space of $\M$ at $\vec{s}_d$, denoted as
$T_{\M}(\vec{s}_d)$. Let $S_\C$ and $S_\T$ be the subsets of
$\RR^{n^2-1}$ corresponding to the Cartan and non-Cartan subspaces $\C$
and $\T$ of $\su{(n)}=\C\oplus\T$.  Since $[-iH_0,\rho_d]=0$ and
$[-iH_1,\rho_d]\in i\T$, we have $A_0s_d=0$ and $A_1s_d\in S_\T$. Let
$H_1=(b_{k\ell})$, and $\vec{v}$ be a column vector consisting of
$\frac{1}{2}n(n-1)$ blocks
\begin{equation}
 \vec{v}^{(k,\ell)} =
\frac{\Delta_{k\ell}}{\sqrt{2}}
 \begin{pmatrix}
 \Im(b_{k\ell}) \\ \Re(b_{k\ell})
\end{pmatrix}.
\end{equation}
Let $B=B_0-\vec{v}\vec{v}^T$ be the restriction of $D_f(\vec{s}_d)$ to
the subspace $S_\T$ as before. Following a similar argument as in
\cite{WS-LyaGeneric} it is easy to see that for $(k,\ell)$ such that
$\Delta_{k\ell}=0$, the eigen-element $(\omega_{k\ell},\vec{e}_{k\ell})$
of $B_0$ is also an eigen-element of $B$ as
$\vec{v}^T\vec{e}_{k\ell}=0$, and that $\vec{e}_{k\ell}$ corresponds to
a direction orthogonal to the tangent space $T_{\M}(\vec{s}_d)$.  The
number of such $(k,\ell)$ is $\bar{N} = 2 \sum_{\ell=1}^k
\binom{n_\ell}{2}$.  We can therefore show that the remaining
eigenvalues of $B$ with eigenvectors corresponding to the directions in
$T_{\M}(\vec{s}_d)$ must have non-zero real parts.  A simple counting
argument shows that the number of these eigenvalues is
$2\binom{n}{2}-\bar n=\dim(\M)$ and thus $\rho_d$ is a hyperbolic point.
Since $\rho_d$ achieves the minimum of $V$, these eigenvalues must have
negative real parts, i.e., $\rho_d$ must be a sink. Hence, there exists
a neighborhood $\N$ of $\rho_d$ such that $\rho(0)\in\N$ will converge
to $\rho_d$ for $t\to+\infty$, which establishes local asymptotic
stability of $\rho_d$.

We can similarly show that any other stationary state $\rho_0$ that is
isolated in $E$ is a hyperbolic fixed point, i.e., all eigenvalues of
linearized system at $\rho_0$ have non-zero real parts.  As $\rho_0$ is
also an unstable critical point of the Lyapunov function $V$ by
Theorem~\ref{thm:degenerate:1}, we can conclude there is an unstable
manifold at $\rho_0$, as in the generic case~\cite{WS-LyaGeneric}.  When
the stationary state $\rho_0$ lies on the critical manifold in $E$ then
the linearized system at $\rho_0$ must again have an unstable manifold;
for otherwise all eigenvalues of $D_f$ at $\rho_0$ would all have
non-positive real parts, and $\rho_0$ would be a local minimum of $V$,
contradicting Theorem~\ref{thm:degenerate:1}.  Therefore, all stationary
states except $\rho_d$ have an unstable manifold of positive dimension,
and the dimensions of all stable manifolds at these unstable stationary
states are less $\dim\M$.  Hence, almost all non-stationary solutions
will not converge to any of these unstable stationary states or their
center manifolds, and therefore must converge to $\rho_d$. Thus,
$\rho_d$ is almost globally asymptotically stable.
\end{IEEEproof}

\begin{example}
With $\rho_d=\diag(\frac{1}{4},\frac{1}{4},\frac{1}{2})$ as in the
previous example we have $n_\M=3^2-2^2-1=4$ and $E$ contains $\rho_d$
and two other stationary states
\(\rho_1=\diag\big(\frac{1}{4},\frac{1}{2},\frac{1}{4}\big)\) and
\(\rho_2=\diag\big(\frac{1}{2},\frac{1}{4},\frac{1}{4}\big)\).  Analysis
of the linearized dynamics shows that the two tangent vectors of the
center manifold at $\rho_\ell$ are also the tangent vectors of $E$.
Therefore, except for $\rho_d$, which is isolated, the points in $E$
form the center manifolds at the stationary states $\rho_\ell$,
$\ell=1,2$.
\end{example}

\begin{figure}
\includegraphics[width=0.9\columnwidth]{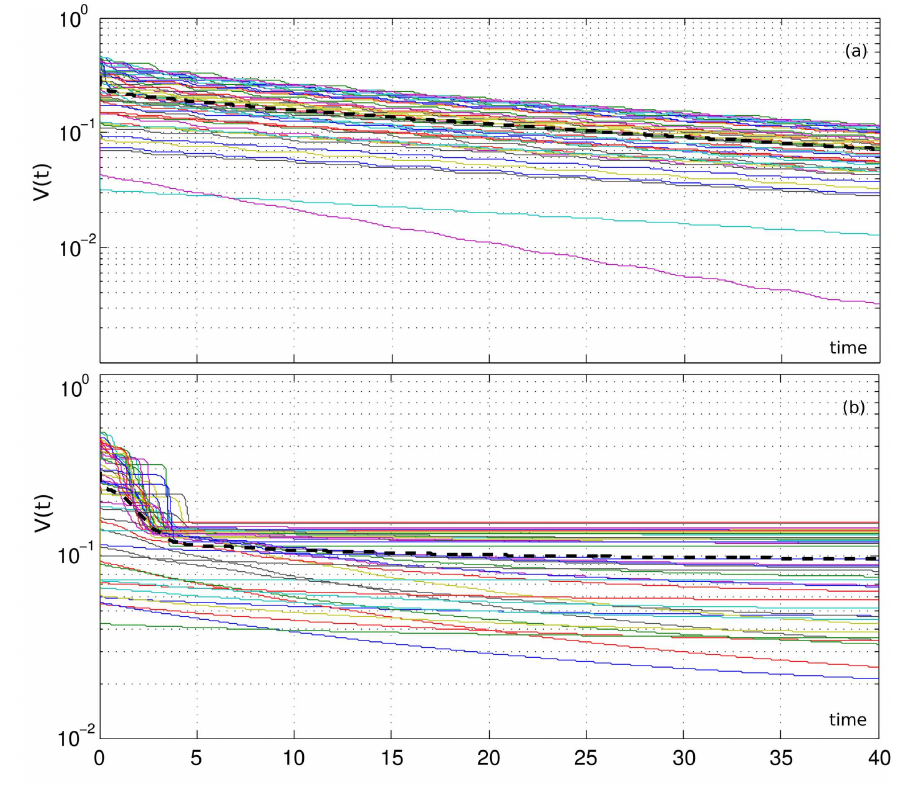}
\caption{Evolution of $V(t)=V(\rho(t),\rho_d(t))$ with $y$-axis in
logarithmic scale.  Each graph shows $V(t)$ for $N=50$ different initial
states $\rho(0)$ for a model system with $H_1$ fully connected and $H_0$
regular but not strongly regular. (a) For $\rho_d=\diag(1,0,0)$ all
trajectories appear linear with negative slope in the semi-logarithmic
plot, indicating exponential convergence; (b) for $\rho_d=\diag(0,1,0)$
most trajectories flat-line after some time at different non-zero
distances, suggesting those trajectories do not converge to $\rho_d$ 
but its center manifold.}
\label{fig:qutrit}
\end{figure}

For generic target states we showed in \cite{WS-LyaGeneric} that for
non-ideal systems the target state itself generally becomes a center on
an attractive center manifold, and the control becomes ineffective.  For
example, when $H_0$ is regular but not strongly regular then the LaSalle
invariant set $E$ usually becomes much larger, forming a center manifold
around the target state $\rho_d$, and most solutions $\rho(t)$ converge
to points on this center manifold other than $\rho_d$.  This is still
true for most non-generic stationary $\rho_d$, for the same reasons, but
unlike in the generic case, there are special target states that are
asymptotically stable even if the Hamiltonian does not satisfy the
criteria for ideal systems.  
\begin{example}
Consider the three-level system with $H_1$ fully connected but $H_0$
not strongly regular
\begin{equation*}
  H_0 = \begin{pmatrix}
         -\omega & 0 & 0 \\
         0 & 0 & 0 \\
         0 & 0 & \omega
         \end{pmatrix}, \quad
  H_1 = \begin{pmatrix}
         0 & 1 & 1 \\
         1 & 0 & 1 \\
         1 & 1 & 0
        \end{pmatrix}.
\end{equation*}
Let $\rho_d=\diag(\alpha_1,\ldots,\alpha_n)$ be the stationary
target state, $\rho=(\beta_{mn})$ and
$\Delta_{k\ell}=\alpha_k-\alpha_\ell$. Then
\begin{equation*}
[\rho,\rho_d]
 = \begin{pmatrix}
0                      & -\beta_{12}\Delta_{12}  & -\beta_{13}\Delta_{23}  \\
\beta_{12}^*\Delta_{12}& 0                       & -\beta_{23}\Delta_{23}  \\
\beta_{13}^*\Delta_{23}& \beta_{23}^*\Delta_{23}& 0
\end{pmatrix},
\end{equation*}
but for the given Hamiltonians $H_0$ and $H_1$, we find that any
point $\rho$ in the LaSalle invariant set satisfies
\begin{equation*}
[\rho,\rho_d]= \begin{pmatrix}
0                      & \gamma  & 0  \\
\gamma^*        &          0    & -\gamma  \\
0                      & -\gamma^*    & 0
\end{pmatrix}.
\end{equation*}
If $\rho_d=\diag(0,1,0)$, then any $\rho\in E$ must satisfy
$\beta_{11} =\beta_{33}$, $\beta_{12} =\beta_{23}$, $|\beta_{13}|
=\beta_{11}$ and $|\beta_{12}| =\beta_{11}-2\beta_{11}^2$. Analogous
to the generic case, the points in LaSalle invariant set form a
center manifold around $\rho_d$ and any solution starting outside
$E$ will stop converging to $\rho_d$ after certain time, as shown in
Fig.~\ref{fig:qutrit} (b).

However, if $\rho_d=\diag(1,0,0)$ then any $\rho\in E-\{\rho_d\}$
satisfies
\begin{equation*}
\rho= \begin{pmatrix}
0                      & 0  & 0  \\
0       &          \beta_{22}    & \beta_{23}\\
0                      & \beta_{23}^*     & \beta_{33}
\end{pmatrix}
\end{equation*}
which corresponds to the maximum value of $V(\rho,\rho_d)$.
Therefore, any solution starting outside $E$ satisfies
$V(\rho(0),\rho_d)<V_{\max}$ and hence converges $\rho_d$, as shown
in Fig.~\ref{fig:qutrit}(a).
\end{example}

This example shows that the non-generic case differs from the generic
one, but for most non-generic target states, a center manifold around
the target state will still appear for non-ideal Hamiltonians, rendering
the control design ineffective.

\section{Pseudo-pure target States}
\label{sec:non-generic:pseudo-pure}

Pseudo-pure states, although a subset of measure zero of all possible
states, deserve special consideration as they include the important
special class of pure quantum states, and non-pure pseudo-pure states
play an important role in applications such as ensemble-based quantum
information processing.  Moreover, for pseudo-pure target states the
analysis simplifies, allowing us to derive the strong stability results
for both stationary and non-stationary target states.

Any pseudo-pure state $\rho_d(0)$ can be written
\begin{equation}
\label{eq:pseudo-pure}
  \rho_d(0)= w \,\Pi_0 + \frac{1-w}{n-1}\,\Pi_0^\perp= w
  \,\Pi_0 + u\,\Pi_0^\perp, \quad 0<w \le 1,
\end{equation}
where $\Pi_0=\ket{\Psi_0}\bra{\Psi_0}$ is a rank-$1$ projector onto
some pure state $\ket{\Psi_0}$, and $\Pi_0^\perp$ is the projector
onto the orthogonal subspace.

\begin{theorem}
\label{thm:pseudo-pure} Assuming ideal Hamiltonian, any solution
$\rho(t)$ such that $V(\rho(0),\rho_d(0))<V_{\rm max}$ converges to
$\rho_d(t)$ as $t\to+\infty$ \emph{except} when $\rho_d(0)=(r_{k\ell})$
has a single pair of non-zero off-diagonal entries of the form
$r_{k\ell}=\frac{1}{2}(w-u)e^{i\alpha}$ and
$r_{kk}=r_{\ell\ell}=\frac{1}{2}(w+u)$.  In the exceptional case any
solution $\rho(t)$ converges to the orbit of $\rho_d(t)$ but in general
$\rho(t)\not\to\rho_d(t)$ as $t\to +\infty$ and $V(\rho,\rho_d)$ can
take any limiting value between $0$ and $V_{\rm max}$.
\end{theorem}

\begin{IEEEproof}
For any $(\rho_1,\rho_2)\in E$, $\rho_1$ and $\rho_2$ must also be
pseudo-pure, with the same spectrum $\{w,u\}$, i.e., $\rho_k=
w\Pi_k+u\Pi_k^\perp$ for $k=1,2$, and we have
\begin{equation}
 \label{eq:pseudo-pure:comm}
 [\rho_1,\rho_2]= (w-u)^2[\Pi_1,\Pi_2]. 
\end{equation}
Thus the LaSalle invariant set contains all points such that
$M=[\Pi_1,\Pi_2]$ is diagonal, according to
Theorem~\ref{thm:lasalle:4}. Let $\Pi_k=\ket{\Psi_k}\bra{\Psi_k}$,
$k=0,1,2$. Setting
\begin{equation}
\begin{split}
 \ket{\Psi_1} &=(a_1e^{i\alpha_1},\ldots,a_ne^{i\alpha_n})^T, \\
 \ket{\Psi_2} &=(b_1e^{i\beta_1},\ldots,b_ne^{i\beta_n})^T,
\end{split}
\end{equation}
we have $M = \ket{\Psi_1}\langle\Psi_1|\Psi_2
      \rangle\bra{\Psi_2}-\ket{\Psi_2}\langle\Psi_2|\Psi_1
      \rangle\bra{\Psi_1}.$
For $(\rho_1,\rho_2)\in E$, the off-diagonal components
of $M$ must vanish:
\begin{equation}
\label{eq:Mkl}
  M_{k\ell} = a_k b_\ell e^{i(\alpha_k-\beta_\ell)} \ip{\Psi_1}{\Psi_2}
             -a_\ell b_k e^{i(\beta_k-\alpha_\ell)}
             \ip{\Psi_2}{\Psi_1}=0,
\end{equation}
for all $k\ne \ell$. Let $\ip{\Psi_1}{\Psi_2}= r e^{i\theta}$. We
have the following two cases.

(a) $r=0$ i.e $\ip{\Psi_1}{\Psi_2}=0$. In this case,
$[\rho_1,\rho_2]=0$, and $V(\rho_1,\rho_2)=V_{\rm max}=(w-u)^2$.

(b) If $r\neq 0$ then (\ref{eq:Mkl}) together with $M_{kk}=0$ gives:
\begin{equation}
 \label{eq:Mk2}.
 a_k b_\ell = a_\ell b_k, \quad
 \beta_k+\beta_\ell = \alpha_k+\alpha_\ell + 2\theta 
\end{equation}
If $a_k=0$ then $0=a_k b_\ell=a_\ell b_k$ for $\ell\neq k$ and we must
have $b_k=0$ as $a_\ell=0$ $\forall\ell$ is not allowed as $\vec{a}$ is
a unit vector.  Ditto for $b_k=0$.  Let $I_+$ be the set of all indices
$k$ so that $a_k,b_k\neq0$.  Then
\begin{equation}
   \frac{a_k}{b_k} = \frac{a_\ell}{b_\ell}, \qquad \forall k,\ell \in I_+
\end{equation}
and thus $\vec{a}=\gamma \vec{b}$, where $\vec{a}=(a_k)$ and
$\vec{b}=(b_k)$. Since $\vec{a}$ and $\vec{b}$ are unit vectors in
$\RR_+^n$, $\gamma=1$ and $\vec{a}=\vec{b}$.

As for the phase equations~(\ref{eq:Mk2}), if $a_k=b_k=0$ then
$M_{k\ell}=0$ is automatically satisfied, thus the only non-trivial
equations are those for $k,\ell\in I_+$.  If the set $I_+$ contains
$n_1>2$ indices then taking pairwise differences of the
$n_1(n_1-1)/2$ non-trivial phase equations and fixing the global
phase of $\ket{\Psi_k}$ by setting $\alpha_{n_1}=\beta_{n_1}=0$
shows that $\vec{\alpha}=\vec{\beta}$.  For example, suppose
$I_+=\{1,2,3\}$ then we have $3$ non-trivial phase equations
\begin{align*}
  \beta_1+\beta_2 & = \alpha_1+\alpha_2 + 2\theta, \\
  \beta_1+\beta_3 & = \alpha_1+\alpha_3 + 2\theta, \\
  \beta_2+\beta_3 & = \alpha_2+\alpha_3 + 2\theta,
\end{align*}
taking pairwise differences gives
\begin{align*}
  \beta_2-\beta_3 & = \alpha_2-\alpha_3,\\
  \beta_1-\beta_3 & = \alpha_1-\alpha_3,\\
  \beta_1-\beta_2 & = \alpha_1-\alpha_2,
\end{align*}
and setting $\alpha_3=\beta_3=0$ shows that we must have
$\alpha_2=\beta_2$ and $\alpha_3=\beta_3$.  Thus, together with
$\vec{a}=\vec{b}$ we have $\rho_1=\rho_2$.  If $I_+$ contains only a
single element then $\ket{\Psi_1}$ and $\ket{\Psi_2}$ differ at most
by a global phase and again $\rho_1=\rho_2$ follows. Incidentally,
note that for $\ket{\Psi_1}=\ket{\Psi_2}$ we have
$\ip{\Psi_1}{\Psi_2}=1$, i.e., $r=1$, $\theta=0$.

The only exceptional case arises when $I_+$ contains exactly two
elements, say $\{1,2\}$, as in this case there is only a single
phase equation $\beta_1+\beta_2=\alpha_1+\alpha_2+2\theta$, and thus
even fixing the global phase by setting $\alpha_2=\beta_2=0$, only
yields $\beta_1-\alpha_1=2\theta$.  This combined with
$\vec{a}=\vec{b}$ gives
\begin{gather*}
  r e^{i\theta} = \ip{\Psi_1}{\Psi_2}
                = a_1^2 e^{2i\theta} + a_2^2
\end{gather*}
and thus $a_1^2 e^{i\theta} + a_2^2 e^{-i\theta} = r$ or
$2i\sin\theta (a_1^2-a_2^2) = 0$. Therefore, either $\theta=0$ or
$a_1=a_2$.  If $\theta=0$ then $\vec{\alpha}=\vec{\beta}$ and
$\rho_1=\rho_2$, which is one possible solution in $E$. If
$\theta\neq 0$, then any $(\rho_1,\rho_2)$ satisfying
\begin{subequations}
\label{eqn:pseudo}
\begin{align}
\ket{\Psi_1}&=2^{-1/2}(1,e^{i\alpha},0,\ldots,0)^T\\
\ket{\Psi_2}&=2^{-1/2}(1,e^{i\beta},0,\ldots,0)^T
\end{align}
\end{subequations}
with $\beta-\alpha=2\theta$ is also in $E$.  Hence, if $\ket{\Psi_0}$
has only two nonzero components with equal norm, e.g., if
\begin{equation}
  \label{eq:rho_limit:t}
  \rho_d(0) =
  \begin{pmatrix} r_{11} & r_{12}& 0 & \ldots & 0\\
                  r_{12}^\dagger & r_{11} & 0 & \ldots & 0\\
                  0        & 0      & u & \\
                  \vdots   & \vdots &        & \ddots \\
                  0        & 0      &        &        & u
  \end{pmatrix},
\end{equation}
with $r_{11}=\frac{1}{2}(w+u)$,
$r_{12}=\frac{1}{2}(w-u)e^{i\alpha}$, and
$\ket{\Psi_0}=2^{1/2}(1,e^{i\alpha},0,\ldots,0)^T$, then $E$
contains all points $(\rho_1,\rho_2)$ satisfying~(\ref{eqn:pseudo}),
which includes $\rho_1=\rho_2$ and $\rho_1\perp\rho_2$.  Since
$\rho_1$ and $\rho_2$ lie on the orbit of $\rho_d(0)$, any solution
$\rho(t)$ will converge to this orbit but we \emph{cannot} guarantee
$\rho(t)\to\rho_d(t)$ as $t\to +\infty$.  For all other $\rho_d(0)$
$E$ contains only points with either $\rho_1=\rho_2$ or
$\rho_1\perp\rho_2$, corresponding to $V=0$ and $V=V_{\rm max}$,
respectively, and since $V$ is non-increasing, any solution
$\rho(t)$ with $V(\rho(0),\rho_d(0))<V_{\rm max}$ will converge to
$\rho_d(t)$ as $t\to +\infty$.
\end{IEEEproof}

\section{Applications for two-qubit systems}
\label{sec:applications}

\begin{figure}
\includegraphics[width=0.9\columnwidth]{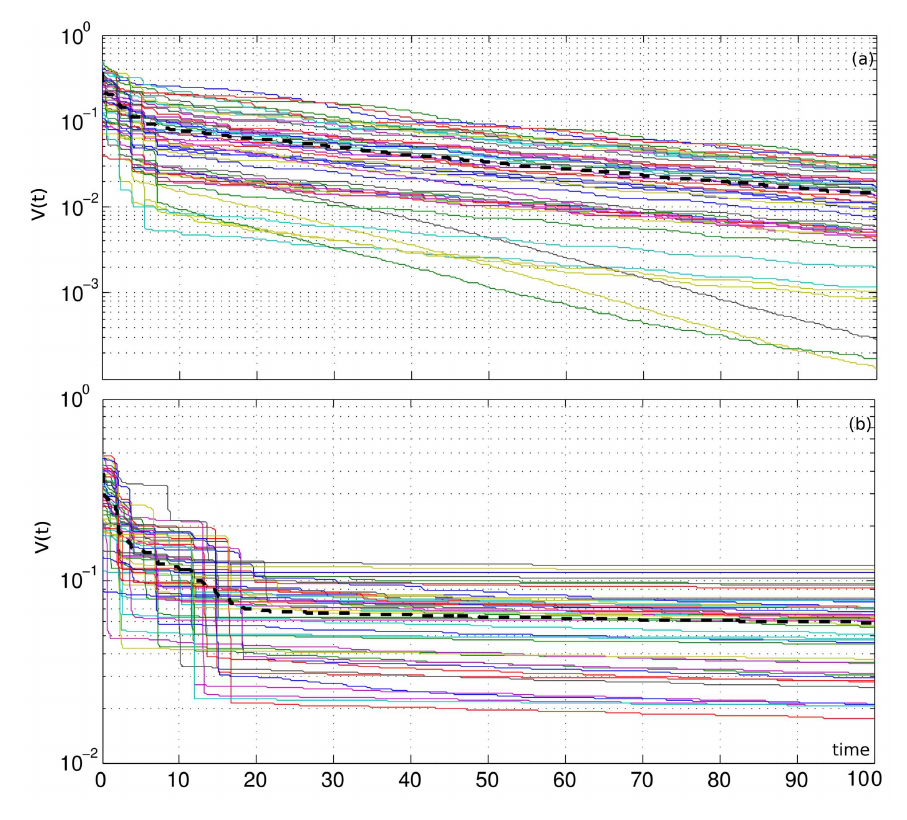}
\caption{$V(t)=V(\rho(t),\rho_d(t))$ with $y$-axis in logarithmic scale.
Each graph shows $V(t)$ for $N=50$ different initial states $\rho(0)$
for a two-qubit system with ideal Hamiltonian. (a) For the target state
$\ket{\Psi}=\frac{1}{\sqrt{5}}(1,0,0,2)^T$ all trajectories converge
exponentially to the target.  (b) For the target state
$\ket{\Psi}=\frac{1}{\sqrt{2}}(1,0,0,1)^T$, most trajectories stop
converging after some finite time and $\lim_{t\to\infty}V(t)>0$.}
\label{fig:qubit2}
\end{figure}

Four-level (or two-qubit) systems are interesting for many reasons,
but for our purposes four-level systems are of interest as $n=4$ is
the lowest dimension that admits states that are neither generic nor
pseudo-pure.  Any stationary state $\rho_d$ has one of the following
forms:
\begin{itemize}
\item generic: $\rho_d=\diag(a,b,c,d)$
\item pseudo-pure: $\rho_d=\diag(a,b,b,b)$
\item other: $\rho_d=\diag(a,a,b,b)$
\end{itemize}
where $a,b,c,d$ are distinct.  The first two cases correspond to the
cases discussed above and in~\cite{WS-LyaGeneric}.  In the third
case, $\rho_d=\diag(a,a,b,b)$, the state manifold $\M$ is
homeomorphic to the flag manifold $\UU(4)/\UU(2)\times\UU(2)$, or
$\SU(4)/\{\exp(\sigma): \sigma\in\C\oplus\T_{12}\oplus\T_{34}\}$ and
has real dimension $n_\M=8$.  Let $(a,a,b,b)$ represent
$\diag(a,a,b,b)$, etc.  There are six stationary solutions:
$(a,a,b,b)$, $(a,b,a,b)$, $(a,b,b,a)$, $(b,a,a,b)$, $(b,a,b,a)$ and
$(b,b,a,a)$. According to the results in
Section~\ref{sec:non-generic:general}, the critical points of
$V(\rho)=V(\rho,\rho_d)$ consist of six components: two isolated
hyperbolic critical points, $(a,a,b,b)$ and $(b,b,a,a)$,
corresponding to the global minimum $V=0$ and maximum $V=2(a-b)^2$,
respectively, and four critical manifolds homeomorphic to $S^2\times
S^2$ on which $V$ assumes the intermediate value $(a-b)^2$.

Calculating the $D_f(\vec{s}_0)$ restricted to $S_\T$ from the
linearized dynamics~(\ref{eqn:linear}) at each stationary state
$\rho_0$, as in the proof of Theorem~\ref{thm:dengerate:2}, shows that
for ideal systems, the dimension of the stable manifold at
$\rho_d=\diag(a,a,b,b)$ is $n_s=8=n_\M$, while at $(b,b,a,a)$, the
dimension the unstable manifold $n_u=8=n_\M$.  Thus these two points are
a hyperbolic sink and source, respectively, with $V_{\min}=0$ and
$V_{\max}=2(a-b)^2$.  At any of the other four stationary states, i.e.,
$(a,b,a,b)$, $(a,b,b,a)$, $(b,a,b,a)$ and $(b,b,a,a)$, we have
$V_0=(a-b)^2$, a stable and unstable manifold of dimension $2$ and two
pairs of purely imaginary eigenvalues, showing that these intermediate
critical points are centers, but the dimension of the corresponding
center manifolds is less than $n_\M$, and each of these centers has an
unstable manifold.  Therefore, the center manifolds and the set of
states that are attracted to them, form a set of measure zero in the
state space, and even in the vicinity of a center manifold, many states
will converge to $\rho_d$.  Thus we can say that $\rho_d$ is almost
globally asymptotically stable.

While this suggests that the Lyapunov design is mostly effective, at
least for ideal Hamiltonian, unfortunately, it still fails for some of
the most interesting problems.  As an example, consider the problem of
Bell state generation, i.e., tracking the trajectory of a maximally
entangled state such as $\ket{\Psi_+}=\frac{1}{\sqrt{2}}
(\ket{00}+\ket{11})$, which is important in quantum information.
Setting $\rho_d(0)=\ket{\Psi_+}\bra{\Psi_+}$ we see immediately that we
cannot track this state because $\rho_d(0)$ satisfies the conditions of
Theorem~\ref{thm:pseudo-pure} for which the method fails: Since $\rho_d(0)$
is pure we have $w=1$ and $u=0$ and taking the standard Pauli product
basis we have $\rho_d(0)=(r_{k\ell})$ with
$r_{11}=r_{14}=r_{41}=r_{44}=\frac{1}{2}$ and $r_{k\ell}=0$
otherwise. Fig.~\ref{fig:qubit2}(b) shows that most initial states
converge to points in the invariant set $E$ that are finite distance
from $\rho_d$. We could console ourselves that the method does work well
for most other target states such as the example in
Fig.~\ref{fig:qubit2}(a).  However, the effectiveness of the method in
these cases is predicated on the assumption of ideal Hamiltonians, which
excludes most typical physical Hamiltonians such as $H_0=0.1\sz \otimes
\sz$ and $H_1=\sx\otimes\ONE +0.9\,\ONE\otimes\sx$ for a system with
constant Ising coupling and local $x$-rotations.  In this case $H_0$ is
not even regular and many of the off-diagonal elements of $H_1$ are
zero, and the invariant set $E$ becomes huge, rendering the method
ineffective.

\section{Conclusion}

We have extended our previous work on Lyapunov control for generic
quantum states to the non-generic case.  Although the LaSalle invariant
set is now much larger, containing critical manifolds, we find that for
ideal Hamiltonian, the stationary target state is still almost globally
asymptotically stable.  For non-ideal systems we find that in general
the target state changes from an isolated asymptotically stable minimum
to a center on a center manifold as in the generic case, but unlike in
the generic case, there are special cases where a non-generic target
state remains asymptotically stable under non-ideal Hamiltonian, though
these states are exceptional.  For the important class of pseudo-pure
states, necessary and sufficient conditions for convergence to the
target state for both stationary and non-stationary cases are derived by
alternative mean.  Finally, application to two-qubit systems shows that
even for ideal systems the small set of (non-stationary) states that are
non-trackable using Lyapunov control includes some of the most
interesting cases such as CAT states
%$\ket{\Psi}=\frac{1}{\sqrt{2}}(\ket{0}+e^{i\phi}\ket{1})$, as well
and maximally-entangled Bell states.
%$\ket{\Psi}=\frac{1}{\sqrt{2}}(\ket{00}+e^{i\phi}\ket{11})$.

\section*{Acknowledgments}

XW is supported by the Cambridge Overseas Trust.  SGS is acknowledges
funding from an \mbox{EPSRC} Advanced Research Fellowship, Hitachi, EU
Knowledge Transfer Programme MTDK-CT-2004-509223 and NSF Grant
PHY05-51164.

\appendix

\noindent A standard basis for the Lie algebra $\su(n)$ is given by
$\{\lambda_{k\ell},\bar{\lambda}_{k\ell},\lambda_k\}$ for $1\le
k<\ell\le n$, where
\begin{subequations}
\label{eq:lambda}
\begin{align}
 &\lambda_k            =  i(\hat{e}_{kk} - \hat{e}_{k+1,k+1}) \\
 &\lambda_{k\ell}      =  i(\hat{e}_{k\ell}+\hat{e}_{\ell k}), \quad
 \bar{\lambda}_{k\ell}=   (\hat{e}_{k\ell}-\hat{e}_{\ell k})
\end{align}
\end{subequations}
and the $(k,\ell)^{\rm th}$ entry of the matrix $\hat{e}_{mn}$ is
$\delta_{km}\delta_{\ell n}$, and $i=\sqrt{-1}$. We have the useful
identities
\begin{subequations}
 \label{eq:lambda_prod}
\begin{align}
  &\Tr(\lambda_{k\ell}\lambda_{k'\ell'})
  = \Tr(\bar\lambda_{k\ell}\bar\lambda_{k'\ell'})
  = -2\delta_{kk'}\delta_{\ell\ell'} \\
  &\Tr(\lambda_{k\ell}\bar\lambda_{k'\ell'})=0
\end{align}
\end{subequations}
and for a diagonal matrix $D=\sum_{k=1}^n d_k\hat{e}_{kk}$ we have
$[D,\lambda_k]= 0$,
\begin{equation}
\label{eq:lambda_comm}
  [D,\lambda_{k\ell}]       = +i(d_k-d_\ell) \bar{\lambda}_{k\ell}, \quad
  [D,\bar{\lambda}_{k\ell}] = -i(d_k-d_\ell) \lambda_{k\ell}.
\end{equation}
The basis~(\ref{eq:lambda}) is not orthonormal but we can define an
equivalent orthonormal basis by normalizing the $n^2-n$ non-Cartan
generators $\frac{1}{\sqrt{2}}\lambda_{k\ell}$ and
$\frac{1}{\sqrt{2}}\bar{\lambda}_{k\ell}$, and defining the $n-1$
orthonormal generators for the Cartan subalgebra \( \sigma_{n^2-n+r} = i
[r(r+1)]^{-1/2} \left(\sum_{s=1}^{r} \hat{e}_{ss} -r\hat{e}_{r+1,r+1}
\right) \) for $r=1,\ldots,n-1$.


\begin{thebibliography}{99}

\bibitem{Vettori}
P. Vettori, ``On the convergence of a feedback control strategy for
multilevel quantum systems,'' in \textit{Proc.} MTNS 2002, 21350.

\bibitem{Ferrante}
A. Ferrante, M. Pavon, and G. Raccanelli, ``Driving the propagator of
a spin system: a feedback approach,'' in \textit{Proc.} 41st IEEE
CDC 2002, 46-50.

\bibitem{Grivopoulos}
S. Grivopoulos and B. Bamieh, ``Lyapunov-based control of quantum
systems,'' In \textit{Proc.} 42nd IEEE CDC 2003, 431-438.

\bibitem{Mirrahimi2004a}
M. Mirrahimi and P. Rouchon, ``Trajectory generation for quantum
systems based on Lyapunov techniques,'' In \textit{Proc.} 
NOLCOS 2004, 291.

\bibitem{Mirrahimi2004b}
M. Mirrahimi and P. Rouchon, ``Trajectory tracking for quantum
systems: A Lyapunov approach,'' In \textit{Proc.} MTNS 2004.

\bibitem{Mirrahimi2005}
M. Mirrahimi, P. Rouchon and G. Turinici, ``Lyapunov control of
bilinear Schrodinger equations,'' \textit{Automatica}, vol.~41, 
pp.~1987-1994, 2005.

\bibitem{Mirrahimi-implicit}
K. Beauchard, J.-M. Coron, M. Mirrahimi and P. Rouchon, ``Implicit
Lyapunov control of finite dimensional Schrodinger equations,'' 
\textit{System Control Lett.}, vol.~56, pp.~388-395, 2007.

\bibitem{Mirrahimi-stabilize}
M. Mirrahimi and R. Van Handel, ``Stabilizing feedback controls for
quantum systems,'' \textit{SIAM J. Control Optim.}, vol.~46, %no.~2,
pp.~445-467, 2007.

\bibitem{altafini1}
C. Altafini, ``Feedback control of spin systems,'' 
\textit{Quantum Information Processing}, vol.~6, pp.~9-36, 2007

\bibitem{altafini2}
C. Altafini, ``Feedback stabilization of quantum ensembles: a global
convergence analysis on complex flag manifolds,'' 
\textit{IEEE Trans. Autom. Control}, vol.~52, pp.~2019-2031, 2007.

\bibitem{D'Alessandro}
D. D'Alessandro, ``Introduction to Quantum Control and Dynamics,'' 
CRC Press, 2007.

\bibitem{WS-LyaGeneric}
X. Wang and S. G. Schirmer, ``Analysis of Lyapunov Method for Control
of Quantum States,'' IEEE TAC submitted, 2008.

\bibitem{lasalle}
J. LaSalle and S. Lefschetz, ``Stability by Liapunov's Direct Method
with Applications,'' New York: Academic Press, 1961.

\bibitem{Wu}
R. Wu, H. Rabitz and M. Hsieh, ``Characterization of the critical
submanifolds in the quantum ensemble control landscapes,'' 
\textit{J. Phys. A}, vol.~41, 015006, 2008.

\end{thebibliography}
\end{document}